\documentstyle[]{article}
\newcommand{\be}{\begin{equation}} \newcommand{\ee}{\end{equation}}
\newcommand{\bea}{\begin{eqnarray}}\newcommand{\eea}{\end{eqnarray}}

\begin{document}
\begin{titlepage}
\title {  Temperature Dependent Nucleon Mass And Entropy Bound Inequality}
\author{M. G. Mustafa$^1$ and A. Ansari$^2$
\\
Institute of Physics, Sachivalaya Marg, \\
Bhubaneswar-751005, INDIA.}
\footnotetext[1]{e-mail: \ mustafa@iopb.ernet.in}
\footnotetext[2]{e-mail: \ ansari@iopb.ernet.in}
\maketitle
\thispagestyle{empty}
\begin{abstract}
Mass of a baryon as a function of temperature is calculated using
colour-singlet partition function for massless quarks (with two
flavours) and abelian gluons confined in a bag with a temperature
dependent bag pressure constant $B(T)$. The non-perturbative aspect
of QCD interaction is included through colour-singlet restriction on
quark-gluon partition function in a phenomenological way. The entropy
bound inequality $ S/E \ \leq \ 2\pi R/\hbar c $, where $S, \ E $ and
$R$ are entropy, energy and radius, respectively of the enclosed
system with $\hbar c \ = \ 197.331 $ MeVfm, is found to be consistent
with the equilibrium solutions of the baryon mass upto a temperature
$T_E$. There is a region of temperature $T_E < T < T_C$ ($T_C$ is
critical temperature for quark-gluon plasma formation) in which no
admissible equilibrium states exist for the bag. We say that the
system expriences a phase jump from hadron to quark-gluon plasma
through thermodynamic non-equlibrium processes.

\end{abstract}
\end{titlepage}
\vfil
\eject

Theory of black hole supplimented with the second law of thermodynamics
could be judiciously  applied to gain new insights into
everyday physics - ordinary thermodynamics, particle physics and
communication theory. The second law of themodynamics states that the
entropy of a closed system tends to a maximum, but this law does not
restrict the value of this maximum. Bekenstein \cite{bek81} suggested
on the basis of black hole thermodynamics \cite{bh75} that if a
physical system can be enclosed in a sphere of radius $R$, then there
is an upper bound on the ratio of the maximum entropy $S$ to its
energy $E$ given by
\bea
 {S\over E} & \ \leq &\ {2\pi R\over {\hbar c}} \nonumber \\
 {\rm or} \ \ \ C \ & = & \ {S\over E} \ - \ {2\pi R \over {\hbar c}} \ \leq
\ 0
\eea

Many objections have been raised \cite{uw82} regarding the original
gedanken experiments involving thermodynamic systems
which led to this conjecture of the bound (1). In early days there
was no rigorous proof; rather over the years it has
become a matter of general belief that the bound (1) is based on the
variety of physical examples supporting it, particularly systems
composed of free quantum fields \cite{bek83} or many quantum mechanical
particles \cite{qk83}. In view of this bound (1) it was also possible
to set limit on the number of generations of quarks \cite{bek82}.
Validity  of this bound in systems with strong gravitional fields
\cite{so81}, with quartic field interaction \cite{bg87} and
superstring systems \cite{bsw86} gives us confidence on the
correctness of this conjecture.
Communication theory \cite{bek82,bek93} is also enriched with the
bound putting limit on the rate at which information can be
transferred in terms of the message energy in a communication channel.
Furthermore, Bekenstein \cite{bek89} has argued that the cosmological
singularity in Friedmann model is inconsistent with bound (1), which
could forbid the appearence of singularity from
thermodynamic view point. A deductive proof of the bound (1) for free
quantum fields enclosed in an arbitary cavity has been given by
Schiffer \cite{sb89,s89}. Very
recently the bound (1) has been used to predict  the limiting
temperature for mesons formation in heavy ion collisions keeping
the meson mass fixed at its bare $T \  = \ 0$ mass in spherical bag
by Dey et al \cite{jm91} and allowing the  meson mass to
depend on temperature in a self-consistent manner in a spherical as
well as a deformed bag by Mustafa et al \cite{ma93}.

Now we would like to test the bound (1) for a nucleonic bag
containing quarks  and abelian gluons with temperature dependent bag
pressure constant $B(T)$ and allowing the nucleon mass to depend on
temperature in a self-consistent manner. Before talking of
calculational details let us first briefly present the outline of the
mathematical steps. The colour-singlet grand canonical partition
function [16-19] for a bag
containing quark, antiquark and gluon is given by

\be
{\cal Z}_{0} \  = \ {\int} \ {d\theta (g)} \ e^{\Omega (\theta )}
\ee

\noindent where $d\theta(g)$ is the invariant Haar measure of colour
$SU(3)$ group  given as
\be
\int \ d\theta(g) \ = \ {1\over{24\pi^2}} \ \int^{+\pi}_{-\pi} \
d\theta_1 \ d\theta_2 \prod^{N_C}_{j < i} \ [2 \sin {1\over 2}
(\theta_j - \theta_i)]^2
\ee

\noindent and

\bea \Omega (\theta ) \ & = & \  \sum_{i} \sum^{\infty}_{K=1}
 {1\over K}\{ [(
N_{C}-1)+2 {\cos K(\theta_{1} - \theta_{2})} \nonumber \\
& & +2 {\cos K(2\theta_{1} + \theta_{2})}+2 {\cos K(\theta_{1}
+ 2\theta_{2})}]{e}^{-K\beta \epsilon^{g}_{i}} \nonumber \\
& &  +(-1)^{K+1} [{\cos K\theta_{1}}+{\cos K\theta_{2}} \nonumber \\
& & +{\cos K (\theta_{1}+\theta_{2})}]
 [{e}^{-K\beta (\epsilon^{q}_{i} - \mu)}
+{e}^{-K\beta (\epsilon^{q}_{i}
+ \mu)}]\} ,
\eea
\noindent where $\theta_i$ is invariant group parameter with
$\sum^{N_{C}}_{i}\theta_{i}  = 0$ , $N_C$ being the number of colour
which is three here. $\epsilon^{q}_{i}$ and $\epsilon^{g}_{i}$
are, respectively, the quark and gluon single particle energies in
the bag, $\mu$ is the chemical potential and $\beta \ = \ {1/T}$, the
inverse of temperature.
Most of the physical quantities can be calculated
using this ${\cal Z}_{0}$. Instead of going to continuum limit all
quantities are computed by performing a finite summation over discrete
states which ensures the finite size of the system [16,17,20].
The colour-singlet thermodynamic potential is given by

\be
\Omega_0 \ = \ \ln ({\cal Z}_{0})
\ee

One can get the net number of quarks ( the excess number of quarks
over antiquark ) by adjusting chemical potential $\mu$ such that
\be
N \ = \ T \ {\partial \Omega_0\over {\partial\mu}} \ = \ N_q \ - \
N_{\bar q} \
\ee
The total energy, $E_T$ and free energy, $F_T$  are, given by
\be
E_T \ = \ T^{2} {{\partial \Omega_{0}}\over{\partial T}} \ + \ \mu
N \ + \  B(T)V \ + \  {d\over R}
\ee
\be
F_T \ = \ -T\Omega_{0} \ + \ \mu N \ + \  B(T)V \ + \  {d\over R}
\ee
\noindent where $V$ is the
volume of the bag and ${d/R}$ accounts for the centre of mass
motion, gluomagnetic interaction and the Casimir energy without any
explicit $T$ dependence. In view of Aerts and Rafelski \cite{ar84} we
have fitted $d \ = \ -3.8274 $ from the nucleon mass $938$ MeV. The
bag pressure $B(T)$ should decrease with increase of temperature.
This was found from QCD sum rule by Dey et al \cite{jm86}, from NJL
model by Li et al \cite{sbb91} and from Soliton bag model by Song et
al \cite{sej92}. We choose the simpler form given in ref.\cite{jm86}
\be B(T) \ = \ B(0) \left [ \ 1 \ - \ \left ({T\over {T_C}} \right
)^4 \ \right ]
\ee
\noindent We use the value of $T_C \ = \ 165$ MeV as obtained in
ref.\cite{ajm90,am92}.
At thermodynamical equlibrium the pressure $P \ = \ - \ {\big (
{\partial F_T \over {\partial V}}\big )}_{T,N}$ is balanced by the
bag pressure which leads to the equilibrium energy $E$ of the bag as

\be
 E \ = \ T^{2} {{\partial \Omega_{0}}\over{\partial T}} \ + \ \mu
N \ + \ {d\over R} \ = \ 3B(T)V
\ee
\noindent and then eq. (7) reduces to
\be
E_T \ = \ M(T) \ = \ 4B(T)V \
\ee

The entropy of the system is given as

\be
S \ = \ - \ {\big (
{\partial F_T \over {\partial T}}\big )}_{V}
\ee

Minimizing the free energy with respect to radius at a given value of
temperature stable solutions of the bag can be obtained. This can be
done graphically plotting $F_T$ vs $R$ as in ref. \cite{ajm90,am92}.
Such a plot is displayed in fig.1 with $B^{1/4}(0) = \ 200$ MeV for
$T \ = 142, \ 143, \ 144, \ {\rm and} \ 144.3 $ MeV. We must mention
that at each point the chemical potential $\mu$ is adjusted such that
the constraint (6) with $N \ = \ 3$ is satisfied. As seen in fig.1
the first three curves have two extremum points; a minimum at a
smaller value of $R$ and a maximum at a larger $R$.  Physically a
meaningful maximum occurs at $R \ = \ 2.5$ fm with $\mu \ \sim \ 0$
for $T$ (say $T_S$)$ \ = \ 130$ MeV (not shown in the fig. 1). The
significance of this value is that for $T \ = \ T_S$ a metastable
solution appears corresponding to the maxima \cite{am92}. If the value of
$T \ < \ T_S$ MeV, there is essentially one extremum (minimum) in
$F_T$, and we have a stable nucleon. Finally at $T_E \ = \ 144.3 $
MeV $>T_S$, the maximum almost disappears and both the extrema meet
in one point. Even for a slightly higher value of $T$ than $T_E$,
there is no equilibrium solutions of the bag implying thermodynamical
instability of the system. This persists in the temperature $T_E < T
< T_C$ because at $T_C$, $B(T) \ = \ 0$ implies that there is no bag
boundary as $R\rightarrow \infty$.

As given by relation (11) the temperature dependent mass $M(T) \ = \
4B(T)V$ at the extremum (with respect to $R$) points of $F_T$.
Equivalently one can also try to match the right hand sides of eq.
(7) and eq.(11) as a function of $R$ keeping in mind that the quark
number constraint (6) has to be always satisfied. In this way we can
calculate $M(T)$ and a corresponding $R(T)$. At such points the bound
inequality (1) is always satisfied. In principle corresponding to
each extremum point of $F_T$ the inequality (1) is satisfied upto
$T=T_E$ and  finally at $T \ > \
T_E \ = \ 144.3 $ MeV the right hand sides of eq.(7) and eq.(11) can
not be matched and, therefore, the value of the $C$ in eq.(1) can not
be computed. Then we may say that
the system has reached a  themodynamic non-equilibrium phase. Thus we
find that the inequality (1) is fully consistent with the equilibrium
solutions of $F_T$ in a variational sense and this $T_E$ is called
the maximum temperature at which baryon can exist in the
thermodynamic equilibrium state. In fig.2 we display a plot of $C(T)$
vs $T$ where $C(T)$ is evaluted at the extremum points of $F_T$ as
given in fig.1. We notice that $C(T)$ is always negative and,
obviously, has two values in the temperature region $T_S \ < \ T \ <
\ T_E$.  $| C(T) |$ larger negative corresponding to the maximum.
This curves reminds us
of the $\mu - T$ plot in ref. \cite{am92} and the branch of $C$ for
$T_S \ < \ T < \ T_E$ corresponds to a metastable region.

The temperature region $T_E < T < T_C$, in which there is no
thermodynamic equilibrium states, corresponds to a non-equilibrium phase
jump of the system. Within the present model two
scenarios are then possible. Suppose that the system is in the state
of thermodynamic equilibrium and the nucleons temperature is raised
to $T_S$ such that it can absorb enough latent heat and move to the
metastable state. Then depending upon the dynamical conditions
temperature reaches  $T_E$ and the system may go to quark-gluon
phase at $T_C$ through non-equilibrium processes.  Or conversely we
may say that heavy ion collisions do not have enough time to
thermalize the hadrons and they can jump to quark-gluon plasma phase
through non-equilibrium processes.

In fig.3 we have shown a plot of $M(T) \ = \ 4B(T)V$ vs $T$. We find
that $M(T)$ increases with the increase of $T$with the rate of
increase being very slow for $T \ < \ 120$ MeV. For $T \ \geq \ T_S$,
$M(T)$ increases very rapidly.  At $T \ = \ T_S \ = \ 130$ MeV there
are two values; one with smaller value of $M(T) \ = \ 1107$ MeV  and
other one with larger value of $M(T) \ = \ 33349$ MeV (not
shown in fig.3), the latter
corresponding to a metastable state.  At $T_E \ = \ 144.3$ MeV two values
coincide to one value of $M(T) \ = \ 1919$ MeV. In finite
temperature field theory \cite{jk} there are two types of masses; one
which appears as the pole of the propagator ( the Green's function )
of the  field, and the other appears in the Yukawa type fall of the
 correlators of the currents. The former is sometimes called the pole mass
and the latter is often known as the screening mass. Of course, both go over
to the unique physical mass in the zero temperature limit.  The
masses measured by Gottlieb et al \cite{g87} are the screening masses at
finite temperature. It is found
that if there is no interaction among the quarks then the self-energy of
the quarks would be screened by the Matsubara frequency $\pi T$. Then
the meson screen masses would go as $2\pi T$ and the
baryon as $3\pi T$ around the critical temperature. We find that in a
simple bag model like picture the baryon mass goes like $4\pi T$ around
$T_E$ ( $144$ MeV ). Thus, in a simple minded bag model description
the dependence of the baryon mass on temperature is rather satisfactory.
Although there is no obvious connections, the obtained $M(T)$ in the
bag model is intriguingly close to that of lattice calculations.

We may point out another outcome of the present calculations. As
seen above, using a temperature dependent bag pressure constant
$B(T)$ with a colour-singlet partition function, we find an
equilibrium temperature $T_E \ \approx \ 145$ MeV when a temperature
independent $B^{1/4}(0) \ = \ 200$ MeV is used. The value of $T_S \ =
\ 130$ MeV for colour-singlet case. If, on the other hand, a colour
{\bf unprojected} calculation is performed, with $B(T)$ using $T_C \
= \ 145$ MeV \cite{am92}, the new value of
$T_S \ = \ 121 \ {\rm MeV \ and} \ T_E \ = \ 125$ MeV. Thus, $T_E \ -
\ T_S \ = \ 4$ MeV for colour unprojected case and about $15$ MeV for
the colour-singlet case. This marked difference in the temperature
width of metastable states is due to the non-perturbative aspect of
the QCD interaction and is almost similar to as that found in ref.
\cite{am92} when temperature independent $B^{1/4} \ = \ 200$ MeV was
used.

Finally we would like to draw the following conclusions: Using a
temperature independent $B^{1/4} \ = \ 200$ MeV earlier (ref.
\cite{am92}) we had found $T_C \ = \ 165$ MeV. Then using a
temperature dependent $B \ = \ B(0)[1 \ - \ (T/T_C)^4]$ with
$B^{1/4}(0) \ = \ 200$ MeV and $T_C \ = \ 165$ MeV a equilibrium
temperature $T_E \approx \ 145$ MeV is found. The temperature window of
metastable states $T_E \ - \ T_S$ is even now (with $B(T)$) about
$15$ MeV for colour-singlet case and about $4$ MeV for the colour
unprojected case.

For the temperature range $T_S$ to $T_E$ the baryon mass increases
rapidly with the increase of $T$, the dependence being roughly as
$4\pi T$. This is qualitatively consistent with the so called screening
mass \cite{g87}.

The entropy bound inequality $C\ < \ 0$ ( eq.1 ) is satisfied at all
extremum points of $F_T$ as a function of $R$. $C$ does not become
positive but it is close to zero ( $| C |$ smallest ) at $T \ \sim \
140$ MeV. For $T \ > \ T_E$ the eqs. (7) and (11) can not be
solved self-consistently except at $R \ \rightarrow \ \infty$
and, therefore, $C$ can not be computed. This  implies that
hadrons can exist in thermodynamic equilibrium states only upto a
temperature $T_E$. Thus, this inequality can certainly be used as a
criterion for testing a system to be bound.

 \vfil
 \eject

\newpage
\begin{figure}[p]

FIGURE CAPTIONS

\caption{Variation of the colour projected free energy, $F_T$ as
a function of bag radius, $R$.}

\caption{Variation of entropy bound inequality, $C$ at the extremum
points of $F_T$, with temperature, $T$.}

\caption{Variation of self-consistently obtained temperature dependent
nucleon mass, $M(T)$ as a function of temperature, $T$.}
\end{figure}
\end{document}